\documentclass[a4paper,aps,pre,twocolumn,groupedaddress,preprintnumbers,floatfix,nofootinbib,showpacs,showkeys]{revtex4}

\usepackage{amsmath,amssymb,bbold}
\usepackage{graphicx,subfigure}
\usepackage[USenglish]{babel}
\usepackage{enumerate}

\newcommand{\bb}{}

\begin{document}

\title{Improving stochastic estimates with inference methods: \\ calculating matrix diagonals}
\author{Marco Selig}
\author{Niels Oppermann}
\author{Torsten A. En{\ss}lin}
\affiliation{Max-Planck-Institut f\"ur Astrophysik, Karl-Schwarzschild-Stra{\ss}e 1, 85741 Garching, Germany}
\received{2011 August 02}
\accepted{2012 February 23}

\begin{abstract}
Estimating the diagonal entries of a matrix, that is not directly accessible but only available as a linear operator in the form of a computer routine, is a common necessity in many computational applications, especially in image reconstruction and statistical inference. Here, methods of statistical inference are used to improve the accuracy or the computational costs of matrix \emph{probing} methods to estimate matrix diagonals. In particular, the generalized Wiener filter methodology, as developed within information field theory, is shown to significantly improve estimates based on only a few sampling probes, in cases in which some form of continuity of the solution can be assumed. The strength, length scale, and precise functional form of the exploited autocorrelation function of the matrix diagonal is determined from the probes themselves. The developed algorithm is successfully applied to mock and real world problems. These performance tests show that, in situations where a matrix diagonal has to be calculated from only a small number of computationally expensive probes, a speedup by a factor of 2 to 10 is possible with the proposed method.
\end{abstract}

\maketitle

\section{Introduction}

Imagine the inside of a black box, which computes an output in a concealed way from a given input, needs to be investigated.
This is a frequent computational task in image reconstruction problems and it might also be necessary in reverse engineering the functionality of a given, already compiled computer code.
In this work, the black box acts as a linear operation on its high-dimensional input, which can be freely chosen in order to probe the box.

However, an internal property of the box that is of interest might be obfuscated and therefore needs to be inferred indirectly by analyzing suitable probes.
In practice, the obfuscation could be just due to the complexity of the box-internal operations, as it happens in image reconstruction problems.
These operations could be any combination of many possible linear operators, like scalar or componentwise multiplications, vector space rotations, Fourier transformations, linear equation solvers,  etc.
The output, however, is assumed to be fully available.

Mathematically, the linear operation performed by the black box can be represented by a matrix in case the input and output spaces are of finite dimension.
In case these spaces have the same dimension, the matrix is a square matrix.
In this paper, we are interested in the diagonal of a matrix for which input and output space can be identified with each other, e.g., in image reconstruction problems by component- or pixelwise correspondence.
There, the matrix diagonal expresses how much a component of the input vector imprints onto the corresponding component of the output vector.
This matrix diagonal, or any other component of the matrix representing the black box operator, can be probed by feeding a number of test vectors into the box and analyzing the results.

The simplest, but most accurate, scheme is to probe any of the diagonal components sequentially, one by one.
This can be done by choosing an input vector that contains only one nonzero component at the location of a certain pixel, and repeating this for all pixels.
The matrix entries can then be read off from the output vectors.
This scheme, however, is computationally very expensive if the dimension of the involved vector spaces is very high.
In image reconstruction, one deals with the space of all possible images, which has the dimension of the number of image pixels.
Probing these sequentially is nearly always computationally prohibitive.

A cheaper, but less accurate, scheme is to feed white noise vectors into the black box and to correlate the outputs to the inputs component- or pixelwise.
Averaging these correlations over a sufficiently large set of input vectors yields an estimator for the desired matrix diagonal.

A first proposal for such a probing scheme can be found in the work by Hutchinson~\citep[and references therein]{H89}. There, the functionality and efficiency of probing for obtaining trace estimates has been proven. Bekas, Kokiopoulou and Saad~\cite{BKS07} extended the probing to Hadamard vectors (rows of the Hadamard matrix) to improve the estimation of diagonals of banded matrices. Those methods have been oriented to applications in density functional theory. Similar problems were approached by Tang and Saad~\citep[and references therein]{TS11} with a focus on nonstochastic estimators. The recent paper by Aune and Simpson~\cite{AS11} transfers the probing technique to the field of information theory, in particular to the calculation of log-likelihoods. Finally, in the extensive work of Rohde and Tsybakov~\cite{RT11}, the noise corrupted observation of unknown matrix entries is investigated from a more mathematical point of view.

In these schemes, the number of required probes is much smaller than in the simple sequential probing scheme.
However, this comes at the price of a lower accuracy of the result due to the stochastic nature of these schemes, which leads to sample variance or uncertainty in the result.
The situation is completely analogous to a noise contaminated signal measurement.
Higher accuracy can be achieved by calculating or measuring more samples, since the noise will average out in the long run.
This, however, might become computationally too expensive at some point.

Noise suppression is a common theme also in the discipline of image and signal reconstruction.
Some \emph{a~priori} knowledge on signal properties may exist, like the presence of spatial smoothness, or the knowledge of limited variance of certain image features (pixels, regions, Fourier modes, etc.). This knowledge can be exploited in order to reduce the impact of the noise on the reconstructed image beyond the simple noise averaging.

Here, we propose to use well-suited image reconstruction techniques, like generalizations of the Wiener filter, for an improved reconstruction of the matrix diagonals.
We show that when determining the diagonals of covariance matrices, as needed in problems of image reconstruction, the usage of suitable filters can improve the accuracy of the inferred matrix diagonal significantly.
The method proposed here yields more accurate results for a given amount of CPU time or, respectively, can reach a given level of accuracy earlier.
In the examined applications, we found a performance increase by a factor of 2 to 10.
This definitely enlarges the number of signal reconstruction problems which can be tackled, and might enable computations which were marginally prohibitive before.

The remainder of this paper is structured as follows.
In Sec.~\ref{sec:prob}, we highlight the importance of obtaining estimates for matrix diagonals, in particular of uncertainty covariance matrices in the context of information field theory (IFT) for signal reconstruction.
The stochastic probing approach is reviewed in Sec.~\ref{sec:freq}.
In Sec.~\ref{sec:bayes}, we propose using filters derived within the framework of IFT to solve the problem of estimating matrix diagonals.
Both methods are verified by simple mock examples as well as by a real-world problem in Sec.~\ref{sec:applications}. We conclude in Sec.~\ref{sec:concl}.

\section{Problem of matrix diagonals}
\label{sec:prob}
   Linear operators are fundamental in any area of computation and thereby often expressed in their matrix representation.

   In the field of information theory, the covariance matrix of a quantity (which equals the inverse of the precision matrix) holds a key role. To stress this, let us consider a multidimensional zero-mean Gaussian,
   \begin{align}
      {\cal G}(\bb{\varphi},\bb{X}) &= \frac{1}{\sqrt{\mathrm{det}\left[ 2 \pi \bb{X}\right]}} \: \exp{\left( - \frac{1}{2} \bb{\varphi}^\intercal \bb{X}^{-1} \bb{\varphi} \right)}
      \text{,}
   \end{align}
   with the covariance matrix $\bb{X} = \left< \bb{\varphi} \bb{\varphi}^\intercal \right>_{{\cal G}} $, where $\bb{\varphi}$ is a random field defined over some pixelized vector space, $\intercal$~denotes a transpose, and $\left< \: \cdot \: \right>_{{\cal G}}$ is the expectation value weighted by this Gaussian. (In this context ``pixel'' is to be understood as a discretized coordinate which elsewhere may be referred to as ``grid point'', ``bin'' or ``voxel''.)

   A diagonal entry of the covariance matrix is the squared standard deviation $\sigma_i$ assigned to pixel $i$ and expresses the pixelwise uncertainty in $\bb{\varphi}$,
   \begin{align}
      \sigma_i^2 &= \left< \varphi_i^2 \right>_{\cal G} = X_{ii}
      \label{standard}
      \text{.}
   \end{align}
   A sophisticated and effective reconstruction tool is the generic filter~\cite{EF10} that we will review in the following. We provide this review in order to further emphasize the importance and problem of obtaining matrix diagonals for stochastic inference. Furthermore, this filter forms also the basis of our proposed algorithm discussed in Sec.~\ref{sec:alg}.

   \subsection{Generic filter}
   \label{sec:cf}
      Signal inference focuses on the reconstruction of some signal $\bb{s}$ in order to explain a set of measurements $\bb{d}$, both of which are connected by a forward data model,
      \begin{align}
         \bb{d} = \bb{R}[\bb{s}] + \bb{n}
         \text{,}
      \end{align}
      where $\bb{n}$ is the noise and $\bb{R}$ a (not necessarily linear) response operator that maps from signal to data space.

      The generalized Wiener filter, as derived, e.g., in Ref.~\cite{EFK08} in a Bayesian framework, is for one thing based on a linear forward data model,
      \begin{align}
         \bb{d} = \bb{R} \: \bb{s} + \bb{n}
         \text{,}
         \label{measurementeq}
      \end{align}
      where the data $\bb{d}$ is a sum of the signal response $\bb{R} \: \bb{s}$ and the noise $\bb{n}$. In this scenario, the response is a linear operator that inherits all aspects of the signal measurement. The generalized Wiener filter arises in case one can assume a Gaussian distribution for the signal's prior and the signal-independent noise in addition to the described forward data model,
      \begin{align}
         P(\bb{s}|\bb{S}) &= {\cal G}(\bb{s},\bb{S})
         \label{s-prior}
         \text{,} \\
         P(\bb{n}|\bb{N}) &= {\cal G}(\bb{n},\bb{N})
         \text{,}
      \end{align}
      where $\bb{S}$ and $\bb{N}$ stand for the signal and noise covariance matrix, respectively. The choice of zero-mean Gaussians shall solely simplify the notation at this point and does not present a restriction for the theory. In fact, choosing an appropriate nonzero mean is often reasonable, as we demonstrate in Sec.~\ref{sec:alg}. In consequence of the forward model, the likelihood of the data given the signal and its response is modeled by the signal-independent noise distribution,\footnote{All noise contributions are marginalized over in an intermediate step, $P(\bb{d}|\bb{s},\bb{R},\bb{N}) = \int {\cal D}\bb{n} \: P(\bb{d}|\bb{s},\bb{n},\bb{R}) P(\bb{n}|\bb{N})$, where $P(\bb{d}|\bb{s},\bb{n},\bb{R}) = \delta(\bb{d} - \bb{R} \: \bb{s} - \bb{n})$ according to Eq.~\eqref{measurementeq}.}
      \begin{align}
         P(\bb{d}|\bb{s},\bb{R},\bb{N}) &= P(\bb{n} = \bb{d} - \bb{R} \: \bb{s}|\bb{s},\bb{R},\bb{N}) \notag \\
         &= {\cal G}(\bb{d} - \bb{R} \: \bb{s},\bb{N})
         \text{.}
      \end{align}
      The actual inverse problem of estimating the signal given the data leads to a posterior that, logically, is also a Gaussian,
      \begin{align}
         P(\bb{s}|\bb{d}) &= {\cal G}(\bb{s} - \bb{m},\bb{D})
         \text{,}
      \end{align}
      with the mean $\bb{m}$ and the covariance $\bb{D}$ that encodes the \emph{a~posteriori} signal uncertainty. The resulting filter formula, whose straightforward derivation is detailed in Refs.~\cite{EFK08,EF10,EW10}, reads
      \begin{align}
         \bb{m} &= \underbrace{\left( \bb{S}^{-1} + \bb{R}^\intercal \bb{N}^{-1} \bb{R} \right)^{-1}}_{\bb{D}} \underbrace{\left( \bb{R}^\intercal \bb{N}^{-1} \bb{d} \right)}_{\bb{j}}
         \label{WF}
         \text{,}
      \end{align}
      where the map $\bb{m}$ is the Bayesian estimator for the signal, i.e.\ its posterior mean, $\bb{D}$ is referred to as \emph{information propagator} and $\bb{j}$ as \emph{information source} in IFT~\cite{EFK08} in close analogy to quantum field theory. Both the signal covariance $\bb{S}$ and the noise covariance $\bb{N}$ needed for this filter are here assumed to be known. A convenient description of these covariances is in terms of the power spectra, the spectra of the eigenvalues of these matrices, as well as the eigenbases in which the covariance matrices are diagonal.

      In the following, we decompose the signal covariance as $\bb{S} = \sum_l C_l \bb{S}_l$, where the $\bb{S}_l$ are projection operators that project onto eigenspaces, the spectral bands (which correspond to the spherical harmonics of equal degree $l$ in our examples in Sec.~\ref{sec:applications}). An analogous decomposition exists for the inverse $\bb{S}^{-1} = \sum_l C_l^{-1} \bb{S}_l^{-1}$, where $\bb{S}_l^{-1}$ is the pseudoinverse of $\bb{S}_l$.

      The signal's power spectrum might be unknown \emph{a~priori}, whereas the eigenbasis can often be guessed from statistical symmetries (e.g., the spherical harmonics basis in case of a statistically isotropic distribution on the sphere). Thus the spectral coefficients $C_l$ allow for a parametrization of the covariance matrix. In such applications without spectral knowledge, the generalized Wiener filter can be extended to a generic filter derived in Ref.~\cite{EF10}. For this purpose, first, a logarithmically flat prior is assumed for the unknown spectral coefficients. Second, the $C_l$-marginalized posterior for the signal is calculated. This posterior then allows for the reidentification of appropriate terms with the spectral coefficients.\footnote{Equation~\eqref{CF} (with $\delta_l = 1$) can alternatively be derived as the maximization of a signal-marginalized posterior with respect to (the logarithm of) the spectral coefficients.} The resulting generic filter formulas are Eq.~\eqref{WF} complemented by a reconstruction rule for the power spectrum, i.e., for each spectral coefficient one calculates
      \begin{align}
         C_l &= \frac{1}{\varrho_l + 2\epsilon_l} \: \mathrm{tr}\left[ \left( \bb{m}\bb{m}^\intercal + \delta_l \bb{D} \right) \bb{S}_l^{-1} \right]
         \label{CF}
         \text{,}
      \end{align}
      where $\varrho_l = \mathrm{tr}\left[ \bb{S}_l^{-1} \bb{S}_l \right]$ are the numbers of degrees of freedom for each spectral band. The parameters $(\delta_l,\epsilon_l)$ characterize the different filter options: two specific forms are the \emph{classical filter}, for which one chooses $(\delta_l,\epsilon_l) = (0,0)$, and the \emph{critical filter}, for which $(\delta_l,\epsilon_l) = (1,0)$. The former can be derived from a ``classical'' maximum \emph{a~posteriori} approximation of the spectral uncertainty marginalized problem. The latter is called ``critical'' because it exhibits (in contrast to the classical filter) only a marginal perception threshold. For a filter with a perception threshold, the signal-to-noise ratio of a spectral mode has to exceed a certain threshold in the data before the filter recognizes it at all. There exists a critical line in the $\delta$-$\epsilon$-plane separating filters that fully suppress bands with insufficient spectral power from filters that do not. The critical filter resides exactly on this line, while the classical filter is in the region with such a perception threshold.

      All in all, Eqs.~\eqref{WF} and \eqref{CF} provide an iterative scheme for the full inverse problem of signal reconstruction with unknown power spectrum, i.e., unknown correlation structure. The signal reconstruction benefits from the additional spectral information recovered from the data since it encodes internal structure of the signal. If, in addition, one lacks the \emph{a~priori} knowledge about the correlation structure of the noise, an analogous approach is conceivable; cf.~\cite{ORE11}.

      The result of the inference is, first, a map $\bb{m}$ describing the posterior mean field of the signal in order to explain the given data $\bb{d}$ and, second, a covariance matrix $\bb{D}$ that encodes the underlying uncertainty correlation structure. The matrix diagonal of $\bb{D}$ is thereby of special importance, since it expresses the pixelwise variance, cf. Eq.~\eqref{standard}, allowing one to make a statement about the uncertainty of the result in each pixel.

      In order to apply the critical or other generic filters we may need to calculate the trace of $\bb{D} \bb{S}_l^{-1}$ in Eq.~\eqref{CF} in each iteration, and we have to evaluate the diagonal of $\bb{D}$ in order to interpret the reliability of our results. This motivates our ambition to develop faster and more accurate matrix probing schemes.

      Generic filters are applied, e.g., in Refs.~\cite{EF10,EW10,ORE11,OJRE11,O11}.

   \subsection{Exact matrix diagonal}
      The diagonal of the uncertainty covariance $\bb{D}$ is a quantity of interest, but unfortunately not directly accessible in most cases. Its calculation involves complex matrix operations, such as matrix inversion; see Eq.~\eqref{WF}. Often, the complete matrix is not known explicitly, only the matrix-vector multiplication is available as a black box in the form of a computer routine which reads in and returns a vector.

      Calculating the diagonal of a matrix $\bb{X}$ of dimension $r$ seems still possible using the canonical basis vectors $\bb{e}^{(k)}$ (with $e_i^{(k)} = \delta_{ik} \: \forall \: i,k \in \{1,\dots,r\}$), since
      \begin{align}
         \mathrm{diag}\left[ \bb{X} \right] &= \sum_k \bb{e}^{(k)} \ast \bb{X} \: \bb{e}^{(k)}
         \label{truediag}
         \text{,}
      \end{align}
      where $\ast$ denotes a componentwise product in the way that $(\bb{a} \ast \bb{b})_i = a_i b_i \: \forall \: i \in \{1,\dots,r\}$.

      It is obvious that this ``true'' diagonal is too expensive computationally because one needs to evaluate the matrix-vector multiplication exactly $r$ times looping through all canonical basis vectors where the dimension $r$ of the problem can be very high ($r \gg 1$). In addition, each of those products alone can be expensive because it may invoke numerical inversion techniques, e.g., a conjugate gradient scheme~\cite{S94}, which is the case in most of the examples in Sec.~\ref{sec:applications}.

\section{Probing estimate}
\label{sec:freq}
   The question arises if one can choose another set of vectors instead of the full set of canonical basis vectors to speed up the computation. Independent and identically distributed random variables stored in a set of vectors $\{\bb{\xi}\}$ (with sample size $|\{\bb{\xi}\}| = A$) work out if they fulfill the property
   \begin{align}
      \left< \xi_i \xi_j \right>_{\{\bb{\xi}\}} &\xrightarrow{A \rightarrow \infty} \delta_{ij}
      \label{conF}
      \text{.}
   \end{align}
   The average $\left< \: \cdot \: \right>_{\{\bb{\xi}\}}$ stands for the arithmetic mean over a set $\{\bb{\xi}\}$ and $\delta_{ij}$ for the Kronecker delta.

   Two of many possible options are (i) equally probable values of $\pm 1$ for the components of $\bb{\xi}$ \cite{H89}\footnote{In Ref.~\cite{BKS07}, a much more sophisticated choice, based on Ref.~\cite{H89}, is presented.} or (ii) zero-mean Gaussian random numbers with unit variance. Both were originally developed for trace estimation. We use (ii) in the following applications.

   Regardless of the choice of the random vectors, the sample average
   \begin{align}
      \left< \bb{\xi} \ast \bb{X} \: \bb{\xi} \right>_{\{\bb{\xi}\}} &\xrightarrow{A \rightarrow \infty} \mathrm{diag}\left[ \bb{X} \right]
      \label{limF}
   \end{align}
   over an infinite set results in the ``true'' diagonal; see the Appendix~\ref{sec:proves}.

   The average over a finite but sufficiently large set ($A < r < \infty$) gives the probing estimate $\bb{f}$ of the matrix diagonal,
   \begin{align}
      \mathrm{diag}\left[ \bb{X} \right] &\approx \left< \bb{\xi} \ast \bb{X} \: \bb{\xi} \right>_{\{\bb{\xi}\}} = \bb{f}
      \label{freqX}
      \text{.}
   \end{align}
   Given this estimator, a trace estimate is obtained by summing up all elements of $\bb{f}$, as
   \begin{align}
      \mathrm{tr}\left[ \bb{X} \right] &\approx \left< \bb{\xi}^\intercal \bb{X} \: \bb{\xi} \right>_{\{\bb{\xi}\}} = \sum_i f_i
      \label{freqtrX}
      \text{.}
   \end{align}
   Since one wants to obtain an estimator in a finite period of time, one has to find an acceptable trade-off between the sample size $A$ and the residual error, where the latter scales with $1/\sqrt{A}$ according to the law of large numbers. Aiming for a certain precision, therefore, requires a particular amount of computation time.

   The estimator given by Eq.~\eqref{freqX} is absolutely generic and applicable to a variety of matrices. Recent applications of it can be found in Refs.~\cite{BKS07,AS11,OJRE11}.

\section{Bayesian estimate}
\label{sec:bayes}

   \subsection{Forward model}
   \label{sec:model}
      Instead of doing a simple averaging of the probes, we now want to develop a Bayesian estimate which exploits additional knowledge of the problem to infer the matrix diagonal from a smaller set of samples. For this purpose, we consider the sampling described by Eq.~\eqref{freqX} as a linear forward model of a measurement process for the signal $\bb{\tilde s} = \mathrm{diag}\left[ \bb{X} \right]$ we are interested in. (In order to avoid confusion, already introduced synonymous quantities that appear now in another context are marked with a tilde.)

      For one sample, $a \in \{1,\dots,A\}$, the measurement equation takes the form
      \begin{align}
         \bb{\tilde d}^{(a)} &= \bb{\xi}^{(a)} \ast \bb{X} \: \bb{\xi}^{(a)} \notag \\
         &= \underbrace{\mathrm{diag}\left[ \big( \xi_1^{(a)} \big)^2,\dots,\big( \xi_r^{(a)} \big)^2 \right]}_{\bb{\tilde R}^{(a)}} \: \bb{\tilde s} + \bb{\tilde n}^{(a)}
         \text{.}
      \end{align}
      For all samples, it is
      \begin{align}
         \bb{\tilde d} &= \left( \bb{\tilde d}^{(1)},\dots,\bb{\tilde d}^{(A)} \right)^\intercal \notag \\
         &= \underbrace{\left( \bb{\tilde R}^{(1)},\cdots,\bb{\tilde R}^{(A)} \right)^\intercal}_{\bb{\tilde R}} \: \bb{\tilde s} + \underbrace{\left( \bb{\tilde n}^{(1)},\dots,\bb{\tilde n}^{(A)} \right)^\intercal}_{\bb{\tilde n}} \notag \\
         &= \bb{\tilde R} \: \bb{\tilde s} + \bb{\tilde n}
         \label{model}
         \text{,}
      \end{align}
      where $\bb{\tilde d}$ represents the ``measured'' data, $\bb{\tilde R}$ the signal response, and $\bb{\tilde n}$ the noise. The contributions from all off-diagonal matrix elements are considered to be noise, i.e.,
      \begin{align}
         \bb{\tilde n}^{(a)} &= \bb{\xi}^{(a)} \ast \left( \bb{X} - \mathrm{diag}\left[ X_{11},\dots,X_{rr} \right] \right) \: \bb{\xi}^{(a)}
         \text{,}
      \end{align}
      and they can be estimated using Eq.~\eqref{model}, once we have an estimator for the signal.

      Note that, if one chooses the random variables $\xi$ to be $\pm 1$, first, one does not have to draw normal variables as originally pointed out by Ref.~\cite{H89} and, second, all the response martices $\bb{\tilde R}^{(a)}$ equal $\mathbb{1}$ and hence do not need to be treated separately for the different samples. This speeds up the algorithm and reduces the memory requirements.

   \subsection{Proposed algorithm}
   \label{sec:alg}
      Our goal is to find an estimator for the matrix diagonal which is close to the minimum mean square error estimate, but still computationally affordable. This estimator has to account for our missing knowledge about the underlying correlation structure. Given these requirements, the generic filter formulas are potentially an appropriate choice. Therefore, our proposed algorithm is based on this filter.

      We start by probing the matrix as described in Sec.~\ref{sec:freq} and as a result obtain a first estimator $\bb{f}$ for our signal, i.e., the matrix diagonal. This additional information changes our state of knowledge about the matrix diagonal in the way that the assumed prior in Eq.~\eqref{s-prior} is not adequate. Although $\bb{f}$ is a sufficient starting value for the iterative part of the scheme, it is not suitable for the construction of a prior, since, after only a few samples, several diagonal entries of the matrix may still be considerably over- or underestimated. By contrast, $\bb{f}$ provides already a sufficiently accurate estimator for the trace; see Eq.~\eqref{freqtrX}. For that reason, we can \emph{a~priori} expect the matrix diagonal $\bb{\tilde s}$ to be distributed around some $\bb{\tilde t}$ rather than around zero, where for all $i \in \{1,\dots,r\}$ we set ${\tilde t}_i = \sum_j f_j / r \approx \left< \mathrm{tr}\left[ \bb{X} \right] \right>_{\{\bb{\xi}\}} / \mathrm{dim}\left[ \bb{X} \right]$. Therefore, the prior of the matrix diagonal is chosen to have a nonzero mean $\bb{\tilde t}$,
      \begin{align}
         P(\bb{\tilde s},\bb{\tilde S}) &= {\cal G}(\bb{\tilde s} - \bb{\tilde t},\bb{\tilde S})
         \text{.}
      \end{align}
      As a consequence, the filter formulas Eqs.~\eqref{WF} and \eqref{CF} undergo a shift,
      \begin{align}
         \bb{\tilde m} &= \bb{\tilde D} \left( \bb{\tilde R}^\intercal \bb{\tilde N}^{-1} \bb{\tilde d} + \bb{\tilde S}^{-1} \bb{\tilde t} \right)
         \label{diagF0}
         \text{,} \\
         \bb{\tilde D} &= \left( \bb{\tilde S}^{-1} + \bb{\tilde R}^\intercal \bb{\tilde N}^{-1} \bb{\tilde R} \right)^{-1}
         \label{diagF1}
         \text{,} \\
         {\tilde C}_l &= \frac{1}{{\tilde \varrho}_l + 2{\tilde \epsilon}_l} \: \mathrm{tr}\left[ \left( \left( \bb{\tilde m} - \bb{\tilde t} \right) \left( \bb{\tilde m} - \bb{\tilde t} \right)^\intercal + {\tilde \delta}_l \bb{\tilde D} \right) \bb{\tilde S}_l^{-1} \right]
         \label{diagF2}
         \text{.}
      \end{align}
      Furthermore, the noise covariance, i.e., its required inverse, is unknown \emph{a~priori} and needs to be estimated for our algorithm. If we use the data model described in Sec.~\ref{sec:model}, $\bb{\tilde N}^{-1}$ can be approximated by the noise given the data and an estimator for the signal,
      \begin{align}
         \bb{\tilde n} &= \bb{\tilde d} - \bb{\tilde R} \: \bb{\tilde m}
         \label{napprox}
         \text{.}
      \end{align}
      We simplify our calculation by using
      \begin{align}
         \bb{\tilde N}^{-1} &= \left( \bb{\tilde n}\bb{\tilde n}^\intercal \right)^{-1} \approx \left( \mathrm{diag}\left[ \bb{\tilde n} \ast \bb{\tilde n} \right] \right)^{-1}
         \label{Napprox}
         \text{.}
      \end{align}
      This is done in order to limit the computational effort, and it can be shown that this corresponds to the treatment of an unknown noise covariance presented in~ Ref.\cite{ORE11} by means of a classical filter.

      Equations~\eqref{diagF0} to \eqref{diagF2} are solved iteratively in the following scheme:
      \begin{enumerate}[(1)]
         \item
            Start with $\bb{\tilde m}^{(\nu=0)} = \bb{f}$ according to Eq.~\eqref{freqX}.
         \item
            Compute $\bb{\tilde n}^{(\nu+1)}$ according to Eq.~\eqref{napprox}.
            \label{step2}
         \item
            Compute ${\tilde C}_l^{(\nu+1)}$ according to Eq.~\eqref{diagF2}, \\ while ignoring $\bb{\tilde t}$ and $\bb{\tilde D}$ for $\nu = 0$.
            \label{step3}
         \item
            Compute $\bb{\tilde m}^{(\nu+1)}$ according to Eq.~\eqref{diagF0} \\ using Eqs.~\eqref{diagF1} and \eqref{Napprox}.
            \label{step4}
         \item
            Repeat steps~(\ref{step2}) to (\ref{step4}) until convergence.
      \end{enumerate}
      As an initial guess for the power spectrum in step~(\ref{step3}), we use an overestimation. This accelerates the convergence process as can be seen in the extreme limits: ${\tilde C}_l \rightarrow \infty: \bb{\tilde m} \sim \bb{\tilde R}^{-1} \bb{\tilde d}$, whereas ${\tilde C}_l \rightarrow 0^+: \bb{\tilde m} \sim \bb{\tilde t}$, i.e., a strong overestimate still gives a nontrivial result for $\bb{\tilde m}$, whereas a strong underestimate gives a nearly trivial one.

      Following Sec.~\ref{sec:cf}, we generally recommend the critical filter, since it does not exhibit a significant perception threshold. Nevertheless, in the presented examples, the correction term $\mathrm{tr}[\bb{\tilde D} \bb{\tilde S}_l^{-1}]$ contributes only marginally to the accuracy and therefore the classical filter, which does not require the calculation of this term, is applied in the following.

\section{Verification \& Application}
\label{sec:applications}

   \subsection{Numerical experiments}
   \label{sec:numexp}
      To verify the proposed algorithm, we perform some numerical experiments that are posed on signals living on the sphere. The examples in this section are represented by all-sky \textsc{HEALPix}~\cite{G05,HEAL} maps with $N_\mathrm{side} = 8$, resulting in $r = 768$ pixels and in a maximal spectral index $l_\mathrm{max} = 23$ of the spherical harmonics basis in which we \emph{a~priori} assume our signal covariance to be diagonal due to a statistical isotropy of the signal.

      The computations were performed using the free open-source mathematics software system \textsc{Sage}~\cite{SAGE} and the \textsc{HEALPix} package.

\begin{figure*}[t]
   \centering
   \subfigure[]
   {  \includegraphics[scale=0.7]{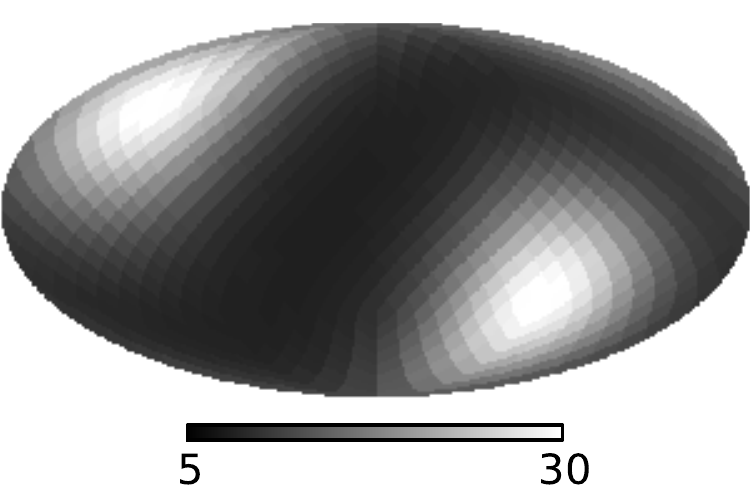}
      \label{fig:trivial_Qt}
   }
   \subfigure[]
   {  \includegraphics[scale=0.7]{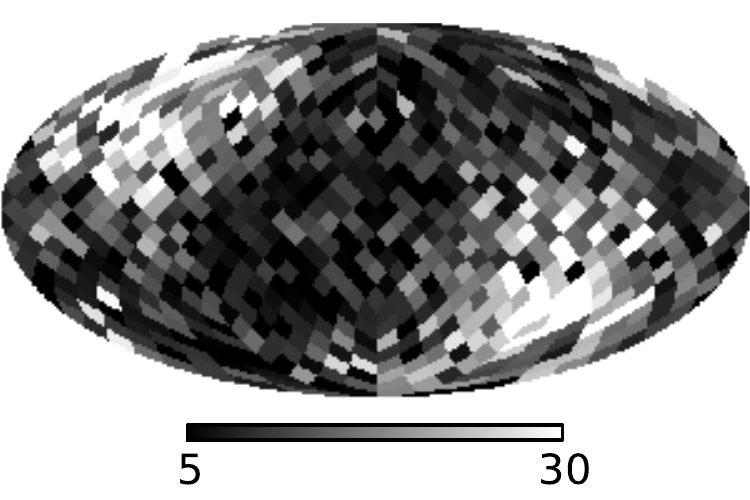}
      \label{fig:trivial_Qf}
   }
   \subfigure[]
   {  \includegraphics[scale=0.7]{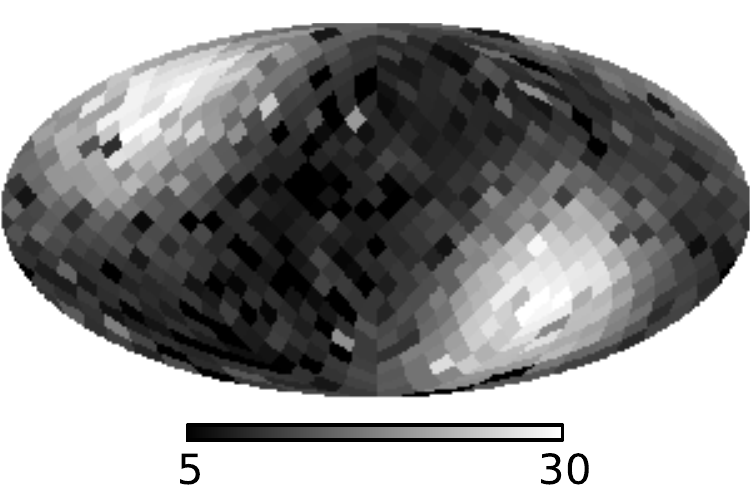}
      \label{fig:trivial_Qm}
   }
   \flushleft
   \caption{Trivial case: \subref{fig:trivial_Qt} the exact matrix diagonal, \subref{fig:trivial_Qf} the probing estimate, and \subref{fig:trivial_Qm} the Bayesian estimate started with four probes, both after around $0.3$~s.}
   \label{fig:trivial_Q}
\end{figure*}
\begin{figure}[t]
   \includegraphics[width=\columnwidth]{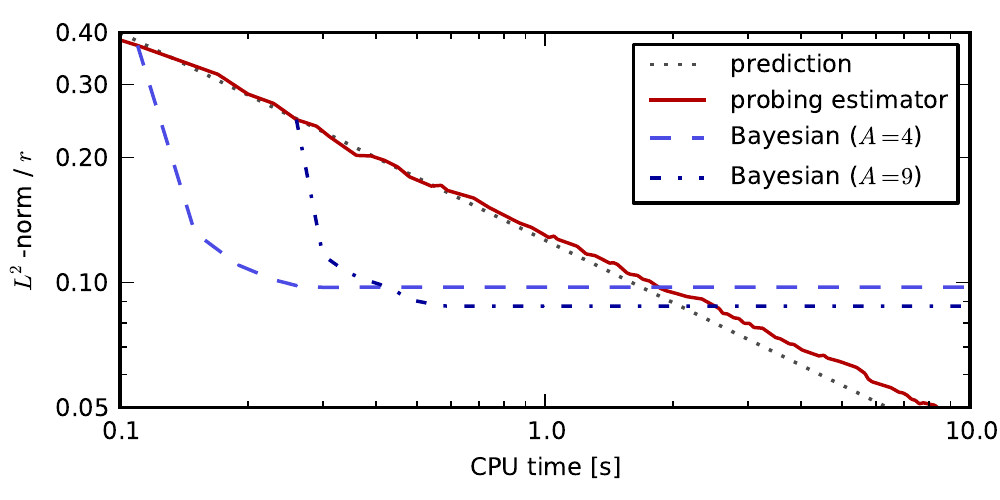}
   \caption{(Color online) $L^2$-norm of the error (divided by the number of pixels $r$) as a function of CPU time for the trivial case: the evolution of the probing estimator (solid), its theoretical prediction $\propto 1/\sqrt{A}$ (dotted) and the Bayesian estimators starting with four (dashed) and nine samples (dashed dotted) are shown.}
   \label{fig:trivial}
\end{figure}

%
      \subsubsection{Trivial case}
         At first, we consider a trivial case where the matrix in question is given explicitly. To ensure that this matrix is a valid covariance matrix, i.e., it is positive and symmetric, we constructed it to be
         \begin{align}
            \bb{X} &= \begin{pmatrix} X_{11} & -1 & & 0 \\ -1 & \ddots & \ddots & \\ & \ddots & \ddots & -1 \\ 0 & & -1 & X_{rr} \end{pmatrix}
            \label{Q}
            \text{,}
         \end{align}
         where the diagonal entries need to fulfill $X_{ii} \geq 2 \: \forall \: i \in \{1,\dots,r\}$ for positive definiteness and are drawn with a simple structure on the sphere; see Fig.~\ref{fig:trivial_Q}.

         The normalized $L^2$-norms of the residual error\footnote{Meaning mathematically $|| \: \mathrm{diag}\left[ \bb{X} \right] - \bb{f} \: ||_2 / r$ or $|| \: \mathrm{diag}\left[ \bb{X} \right] - \bb{\tilde m} \: ||_2 / r$, respectively, to be exact.} serve as an accuracy measure and are shown as a function of CPU time in Fig.~\ref{fig:trivial}.

         Although $\bb{X}$ as an operator could be implemented very efficiently, we use the much more expensive full matrix multiplication to have realistic computational costs like those of applying a more complex matrix. But this trivial case should only hold as a proof of concept; more sensible examples are discussed in the following.

         As one can clearly see in Fig.~\ref{fig:trivial}, the probing estimator improves continuously with an increasing number of probes and shows an overall proportionality to $1/\sqrt{A}$ as argued in Sec.~\ref{sec:freq}. However, the Bayesian estimator given a set of samples converges in only a couple of iterations to a result with an accuracy the pure probing will first reach after investing a factor of a few more CPU time. For a fixed amount of computation time, the Bayesian estimator excels the probing estimator, as can be seen in Fig.~\ref{fig:trivial_Q}. It is also evident that the Bayesian estimator must reach a lower limit in its progress because only a limited data set is provided containing finitely accurate information.

\begin{figure*}[t]
   \centering
   \subfigure[]
   {  \includegraphics[scale=0.7]{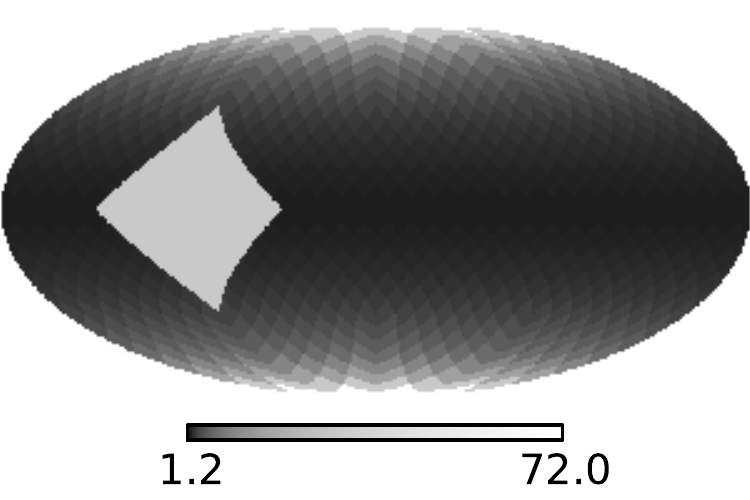}
      \label{fig:XN}
   }
   \subfigure[]
   {  \includegraphics[scale=0.7]{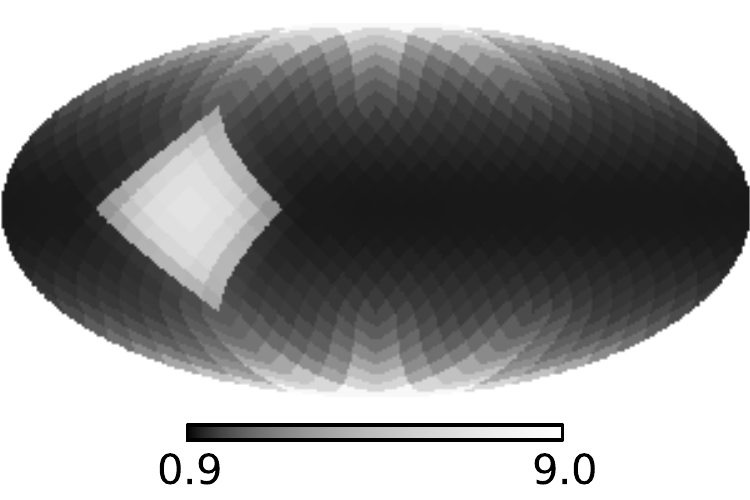}
      \label{fig:Xt}
   }
   \flushleft
   \caption{Realistic case: \subref{fig:XN} the matrix diagonal of the mock noise covariance $\bb{N}$ and \subref{fig:Xt} the ``true'' diagonal of the propagator $\bb{D} = (\bb{S}^{-1}+\bb{N}^{-1})^{-1}$.}
   \label{fig:X_ND}
\end{figure*}
\begin{figure}[t]
   \includegraphics[width=\columnwidth]{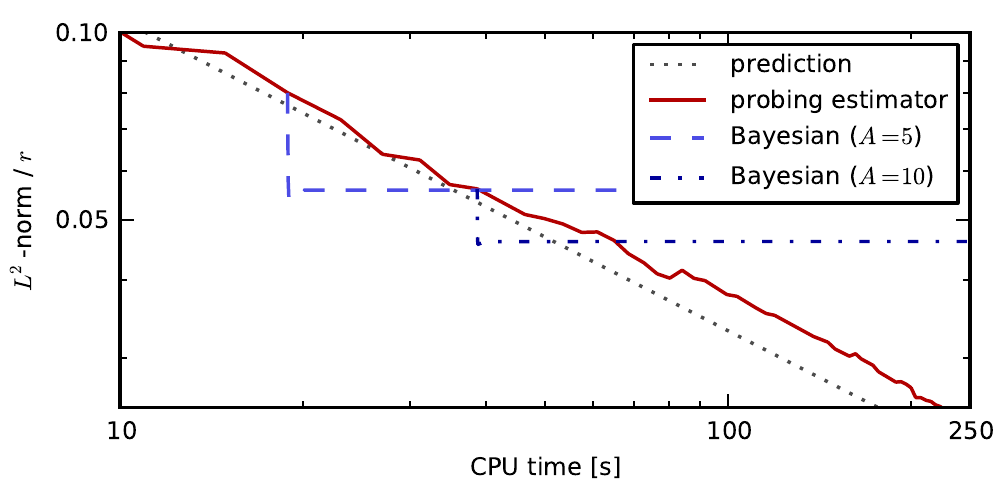}
   \caption{(Color online) Same as Fig.~\ref{fig:trivial}, only for the propagator for the realistic case.}
   \label{fig:realistic}
\end{figure}

%
      \subsubsection{Realistic case}
         For a more realistic mock example, we consider a covariance matrix $\bb{D} = (\bb{S}^{-1}+\bb{N}^{-1})^{-1}$ similar to the one described by Eq.~\eqref{WF} where the signal covariance $\bb{S}$ is completely defined by a power spectrum,
         \begin{align}
            C_l &\propto \left( \mathrm{max}\left\{1,l\right\} \right)^{-2}
            \text{.}
         \end{align}
         The noise covariance $\bb{N}$ is characterized by two effects, first, high noise in one of the twelve \textsc{HEALPix} basis pixels representing a defect in the detector and, second, smoothly increased noise toward the poles imitating an observational effect.\footnote{The noise variance is assigned to each pixel $i$ according to $N_{ii}^{-1} = (0.005 + 8 \left( h_i (h_i - h_\mathrm{max}) \right)^2 / h_\mathrm{max}^4)$, where $h_i$ is the \textsc{HEALPix} ring number associated to the pixel $i$.} The described noise covariance and the resulting propagator $\bb{D}$ are illustrated in Fig.~\ref{fig:X_ND}, where one can see the conservation of the noise structure and the smoothing effect of the power spectrum.

         The performance of both algorithms is shown in Fig.~\ref{fig:realistic}. Our algorithm performs qualitatively in the same way as in the trivial case, but the overall gain in accuracy or time is quantitatively lower. It is also noticeable that the relative advantage of the proposed method decreases with the number of used random vectors. Consequently, the matrix diagonal inference method pays off best in cases where a rough estimate using only a few probes is sufficient.

   \subsection{Faraday sky uncertainty}
   \label{sec:FSU}

\begin{figure*}[t]
   \centering
   \subfigure[]
   {  \includegraphics[scale=0.7]{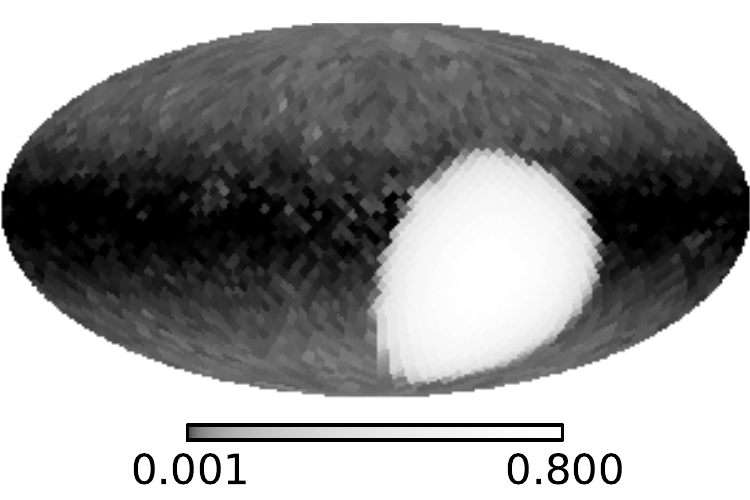}
      \label{fig:Ft}
   }
   \subfigure[]
   {  \includegraphics[scale=0.7]{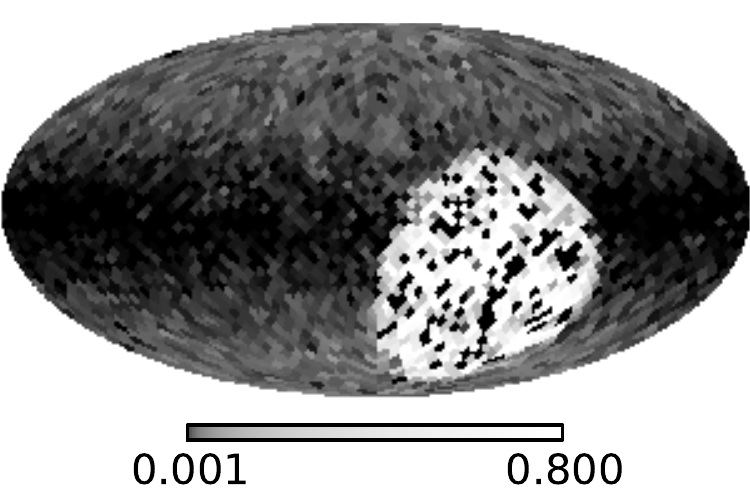}
      \label{fig:Ff}
   }
   \subfigure[]
   {  \includegraphics[scale=0.7]{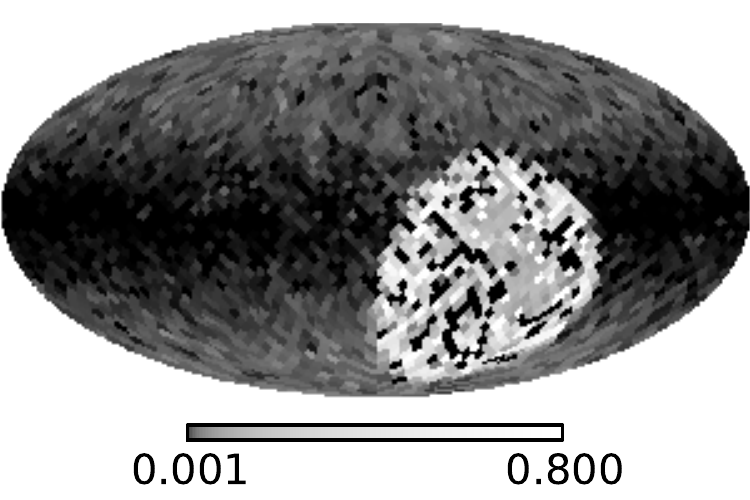}
      \label{fig:Fm}
   }
   \flushleft
   \caption{Diagonal of the propagator for the reconstruction of the galactic Faraday depth. Panel~\subref{fig:Ft} shows the result of the exact calculation according to Eq.~\eqref{truediag}, panel~\subref{fig:Ff} the probing result after ten iterations, and panel~\subref{fig:Fm} the result of the Bayesian estimator, using ten random vectors as well.}
   \label{fig:Faradaymaps}
\end{figure*}
\begin{figure}[t]
   \includegraphics[width=\columnwidth]{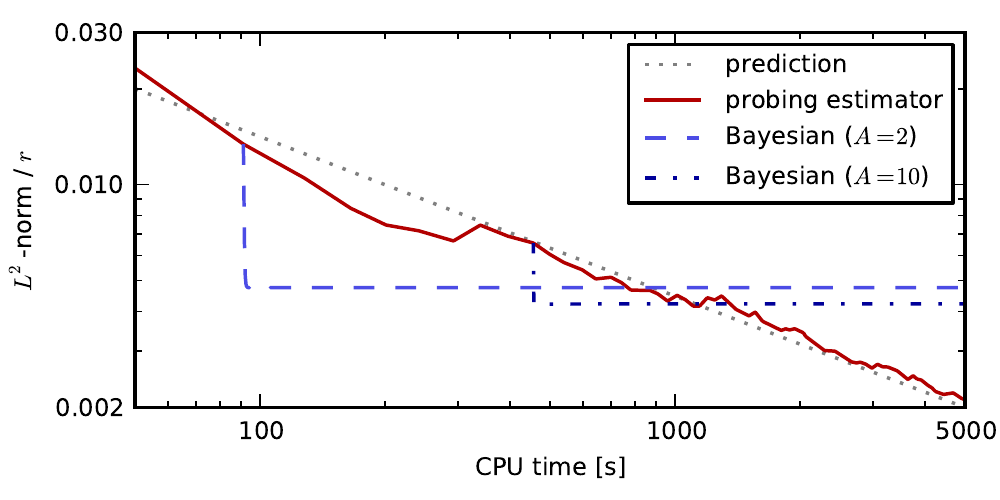}
   \caption{(Color online) Same as Fig.~\ref{fig:trivial}, only for the propagator for the reconstruction of the galactic Faraday depth.}
   \label{fig:Faradaycomp}
\end{figure}

		Next, we attempt to use our algorithm in a real physical application. We consider the inference problem discussed in Ref.~\cite{OJRE11}. In that work, an all-sky signal -- the galactic Faraday depth -- was reconstructed from measurements in $37\,543$ individual directions \cite{TSS08}. The data were modeled according to a linear measurement procedure, Eq.~\eqref{measurementeq}, with a response $\bb{R}$ that is nonzero only for directions in which measurements had been made and is larger within the galactic plane.

		To reconstruct the field, the critical filter algorithm that was discussed in Sec.~\ref{sec:cf} was used, yielding an estimate for the posterior mean $\bb{m}$ of the signal field $\bb{s}$, as well as an estimate for the components of the angular power spectrum of this field, $C_l$. In addition, a map showing the uncertainty of the signal estimate $\bb{m}$, given by $\mathrm{diag}\left[\bb{D}\right]$, is provided in Ref.~\cite{OJRE11}. This was calculated from the information propagator $\bb{D}$, which takes on the form \eqref{diagF1}, by applying the probing estimator discussed in Sec.~\ref{sec:freq}.

		Here, we show how the application of our Bayesian algorithm to this problem can improve the accuracy and speed up the calculation of this matrix diagonal. In order to be able to compare the results of the probing and Bayesian estimators to the correct matrix diagonal, we reduce the dimensionality of the problem to facilitate the exact calculation of the diagonal via Eq.~\eqref{truediag}. We do this by reducing the resolution of the all-sky map with respect to the one presented in Ref.~\cite{OJRE11} to \textsc{HEALPix} parameter $N_\mathrm{side}=16$, leading to $3\,072$ pixels, and truncating the reconstructed power spectrum at $l_\mathrm{max}=47$. Furthermore, we use a coarser version of the response matrix. In this way, a coarse-grained version of the propagator $\bb{D}$ is defined and we can calculate its diagonal exactly, as well as by using the probing estimate and our Bayesian extension.

		Figure~\ref{fig:Faradaymaps} shows the results of these calculations, where the matrix-vector multiplication was conducted for ten different random vectors in the case of the probing and Bayesian estimators. The sphere in this case corresponds to a map of the whole sky, where the disk of the Milky Way extends horizontally in the middle of the image. While still exhibiting a large amount of noise, both the probing and Bayesian results show roughly the right structure after only a few iterations. This structure is determined mainly by an oval region of high uncertainty, i.e.\ large diagonal entries, where no observations had been made and a dependence on galactic latitude due to the larger signal response within the galactic plane.

		From Fig.~\ref{fig:Faradaymaps} alone, it is hard to judge whether the Bayesian estimator leads to an improvement over the probing one. We therefore plot again the $L^2$-norm of the difference between the estimated matrix diagonal and the ``true'' one as a function of CPU time in Fig.~\ref{fig:Faradaycomp}. Shown is the curve for the pure probing estimator as well as two examples for Bayesian improvements, using two and ten random vectors, respectively. It is evident that for both cases the Bayesian method gives a boost in accuracy with only marginal time consumption. The absolute and relative improvement is larger if one uses fewer random vectors. This shows again that the main strength of the Bayesian method does not lie in the absolute accuracy that can be reached, but rather in the speedup it provides for obtaining an estimate for the matrix diagonal with intermediate accuracy.

\section{Conclusions}
\label{sec:concl}

This work aims at transferring successful signal inference methods to the realm of numerical computation.
In particular, the problem of inferring internal properties of a computational black box is analyzed.
To this end, the well known method of stochastic probing of matrix diagonals is augmented by a sophisticated signal reconstruction scheme.
This inference algorithm interprets the probing of the matrix diagonal as a numerical measurement including unwanted uncertainties, or plainly noise.

As in signal reconstruction, additional knowledge on the matrix diagonal, i.e., our signal, can be exploited to improve the accuracy of the obtained result.
We use known symmetries of the underlying continuous structure of the matrix diagonal, namely approximate statistical isotropy.

This improves the estimates acquired from probing, without increasing computational costs significantly.
In order to achieve the same level of accuracy, the traditional probing method requires computational costs, which are a factor of 2 to 10 times larger than the proposed scheme, in the cases investigated here.
This algorithm is especially effective when matrix diagonals need to be calculated only roughly, since the relative gain in accuracy is larger if only a few probes are available.

This has been shown in numerical examples as well as for the uncertainty map appearing in the reconstruction of the galactic Faraday depth.

\section{Acknowledgments}
   We thank Mike Bell, Henrik Junklewitz, Georg Robbers and three anonymous referees for the insightful discussions and productive comments on the manuscript.

\appendix
\section{Proof of the probing estimator}
\label{sec:proves}
   Here we prove that Eq.~\eqref{limF} is indeed implied by Eq.~\eqref{conF}. Given a sufficiently large but not necessarily finite set $\{\bb{\xi}\}$ (with $|\{\bb{\xi}\}| = A$), the condition given by Eq.~\eqref{conF} becomes
   \begin{align}
      \lim_{A \rightarrow \infty} \left< \xi_i \xi_j \right>_{\{\bb{\xi}\}} &= \lim_{A \rightarrow \infty} \frac{1}{A} \sum_{a=1}^A  \xi_i^{(a)} \xi_j^{(a)} = \delta_{ij}
      \text{.}
   \end{align}
   Inserting this equality in Eq.~\eqref{limF}, without loss of generality restricted to the average's component $i \in \{1,\dots,r\}$,
   \begin{align}
      \left( \left< \bb{\xi} \ast \bb{X} \: \bb{\xi} \right>_{\{\bb{\xi}\}} \right)_i &= \frac{1}{A} \sum_{a=1}^A \sum_{j=1}^r \xi_i^{(a)} X_{ij} \xi_j^{(a)} \notag \\
      &= \sum_{j=1}^r X_{ij} \underbrace{\frac{1}{A} \sum_{a=1}^A  \xi_i^{(a)} \xi_j^{(a)}}_{\rightarrow \delta_{ij}}
      &\rightarrow X_{ii}
      \text{,}
   \end{align}
   proves the statement.


\newcommand{\aap}{A\&A}
\bibliography{paper.bib}

\end{document}